\documentstyle[preprint,aps]{revtex}
\begin{document}
\draft

\title{
Universal Statistics of Transport in Disordered Conductors}
\author{ Hyunwoo Lee$^1$, A. Yu. Yakovetz$^2$,
  and L. S. Levitov$^{1,2}$ }
\address{
$^1$Department of Physics, Massachusetts Institute of Technology,\\
     77 Massachusetts Ave., Cambridge, MA 02139,\\
$^2$Landau Institute for Theoretical Physics,\\
     2 Kosygin St., Moscow 117334, Russia}
\maketitle

\begin{abstract}
In low temperature limit, we study electron counting statistics
of a disordered conductor.
We derive an expression for the distribution
of charge transmitted over a finite time interval by using
a result from the random matrix theory of quasi one dimensional
disordered conductors.
In the metallic regime, we find that the peak of the distribution
is Gaussian and shows negligible sample to sample variations.
We also find that the tails of the distribution are neither
Gaussian nor Poisson and exhibit strong sample to sample variations.

\end{abstract}
\narrowtext

The physics of current fluctuations at low temperature presents
an interesting quantum mechanical problem. Classical Johnson-Nyquist noise
formula\cite{Nyquist} gives a good description of current fluctuations
due to thermal fluctuations. However, at low temperature
thermal fluctuations are small and a new type of noise becomes
important. At low temperature, the quantum nature
of the current and the discreteness of electron charge is
the main source of current fluctuations and due to these reasons,
this noise is called ``quantum shot noise''.

Lots of the low temperature current fluctuation studies
deal with a disordered conductor because it has a simple and well
established mathematical description based on Landauer's
approach\cite{Landauer}. Lesovik\cite{Lesovik} and
Yurke and Kochanski\cite{Yurke} studied the quantum shot noise
in a two-terminal conductor using this approach
and found an expression for noise power
which is a factor $1 \! - \! T$ off the classical shot noise,
where $T$ is a transmission coefficient. This analysis was
generalized to a multiterminal conductor by
B\"uttiker\cite{Buttiker}, and he also found the reduction of
the noise. Physically, the noise reduction is due to the Fermi
statistics that leads to correlation of transmission events.

In Landauer's approach, details of current transport are
determined by transmission coefficients
and there have been many works on the distribution of the coefficients.
In the past decade, the random matrix theory
of disordered conductors was developed\cite{Stone}, motivated by
theoretical discovery and experimental observation of the universal
conductance fluctuations. The theory succeeded
in providing a complete characterization of the distribution
and it also succeeded in providing the insight on the origin
of the universal conductance fluctuations.
One of the fundamental results in the random matrix theory
is the universality of the distribution in the metallic
regime and the universality provides a link between
the microscopically calculated transmission coefficients and
the macroscopically measurable conductance.

Application of the universal distribution to
current fluctuations also provides insights on the current
fluctuations of a disordered conductor.
Beenakker and B\"uttiker\cite{Beenakker} calculated the sample
averaged noise power using the universal
distribution and found that it depends only on the conductance
and that it is one-third of the classical value.

Noise power is a good measure of noise magnitude and its
study revealed the reduction of noise due to Fermi correlation.
However, compared to Johnson-Nyquist noise, our understanding
of quantum shot noise is limited and not many things are known
besides the noise magnitude.
Our goal in this paper is to explore the physics of low temperature
current fluctuations beyond the scope of noise power.
For our purpose, it is useful to look at
the behavior of current fluctuations in the time domain, which
brings one to the notion of counting statistics of charge
transmitted in a conductor over fixed time. A previous study
of the counting statistics for a single channel conductor revealed
that the attempts to transmit electrons are highly correlated and
almost periodic in time, which leads to binomial
statistics\cite{LevitovBinomial}.

In the time domain study of current fluctuations,
the charge $Q(t)$ (measured in the units of $e$) transmitted over
a time interval $t$ is the quantity of interest and the probability
distribution $P(Q(t))$ reflects
the nature of the current fluctuations. Even at zero temperature,
$P(Q(t))$ has finite peak width due to the quantum nature of current.
To study $P(Q(t))$, it is useful to introduce the characteristic function
$\chi(\lambda)$,
\begin{equation}
  \chi(\lambda)=\sum_{{\rm integer } \,Q} e^{iQ\lambda } P(Q) \,\,
    \mbox{ for } -\pi< \lambda < \pi \, ,
  \label{eqn:definition}
\end{equation}
because in many cases, $\chi(\lambda)$ is easier to calculate than
$P(Q(t))$ itself. $\chi(\lambda)$ is a Fourier transform of $P(Q(t))$
and so once $\chi(\lambda)$ is obtained, we can either take
an inverse Fourier transform of it to get an explicit
expression for $P(Q(t))$, or  expand it as follows to get cumulants
of the distribution:
\begin{equation}
  \log \chi(\lambda)=\sum_{k=1}^{\infty} { (i\lambda)^k \over k! }
    \langle\langle  Q^k \rangle\rangle.
  \label{eqn:formulacumulant}
\end{equation}

In the linear transport regime, we derive a general expression
for $\chi(\lambda)$ in terms of transmission coefficients and
by combining it with the transmission coefficients distribution
for quasi one dimensional conductors,
we show
\begin{equation}
\overline{ \log \chi(\lambda)}= {GVt \over e} {\rm arcsinh}^2
   \sqrt{e^{i\lambda}-1} \,  \ ,
\label{eqn:result}
\end{equation}
where $V$ is the dc voltage, $G=g(e^2/h)=(Nl/L)(e^2/h)$ is
the average conductance,
and the bar on the left hand side represents the sample average.
Cumulant expansion of Eq.~(\ref{eqn:result}) implies that in average,
for $GVt/e \gg 1$, $P(Q(t))$ has a Gaussian peak at $GVt/e$
with $\langle\langle Q^2(t) \rangle\rangle=GVt/3e$.
It also implies that even though the peak is Gaussian,
the tails show deviation from both Gaussian and Poisson distributions.
We estimate sample to sample variations of $P(Q(t))$
by studying variances of various quantities
and find that for $GVt/e \gg 1$ sample to sample variations of
$P(Q(t))$ appear only in the tails of $P(Q(t))$ and that around the peak,
$P(Q(t))$ is universal.

Before we present the derivation of above result, we stress that
there are two kinds of averages involved. To avoid confusion,
we will use a bar($\overline{\cdots}$) for
an ensemble average, or an average over samples,
and a bracket($\langle \cdots \rangle$) for
a quantum average, or a quantum expectation value.
Also we reserve a double bracket($\langle\langle \cdots \rangle\rangle$)
for a cumulant of a quantum expectation value and ``var''(var$(\cdots)$)
for $\overline{\cdots^2}-{\overline{\cdots}}^2$.

Now, let us derive Eq.~(\ref{eqn:result}). Following
the Landauer's approach\cite{Landauer}, we consider a conductor
sandwiched between two perfect leads. In the linear transport regime,
scattering properties of a conductor are described by a unitary
scattering matrix ${\hat S}$ that relates incoming and outgoing
amplitudes, $I_{L(R)}$ and $O_{L(R)}$:
\begin{equation}
  {\hat S} \left( \begin{array}{c} I_L \\ I_R \end{array} \right)
    = \left( \begin{array}{c} O_L \\ O_R \end{array} \right),
  \label{eqn:scattering}
\end{equation}
where the subscripts $L$ and $R$ stand for the left and the right leads.

The unitarity of ${\hat S}$ is due to the current conservation, and
it allows a system to be decomposed into independent
channels\cite{Mello}. Then the decomposition motivates one to study
single channel transport first, where a transmission coefficient $T$
determines the transport.
Recently, the counting statistics of the single channel transport was
studied\cite{LevitovBinomial}. In the low temperature limit
$(k_B{\cal T} \ll eV)$, the characteristic function
$\chi_1(\lambda)$ of a single channel system becomes
\begin{equation}
  \chi_1(\lambda)=(pe^{i\lambda}+q)^{M},
  \label{eqn:binomial}
\end{equation}
where $p=T$, $q=1-T$, $M=eVt/h$\cite{comment} and
$M \gg 1$ is assumed.
The inverse Fourier transform of Eq.~(\ref{eqn:binomial}) gives
the binomial distribution, which implies that the intervals between
subsequent attempts to transmit electrons are quite regular.
This regularity is due to Pauli exclusion principle.

Having the characteristic function of a single channel, we write
the total characteristic function $\chi(\lambda)$ as a product,
\begin{equation}
  \chi(\lambda)=\prod_j (T_j e^{i\lambda}+1-T_j)^M,
  \label{eqn:product}
\end{equation}
where $T_j$ is a transmission coefficient of channel $j$.
The product form Eq.~(\ref{eqn:product}) follows from
the mutual independence of channels.
By taking logarithm of Eq.~(\ref{eqn:product}), we get
\begin{equation}
  \log \chi(\lambda)=M\sum_j
    \log\Biggl(T_j e^{i\lambda}+1-T_j \Biggr),
  \label{eqn:logchi}
\end{equation}
and by expanding Eq.~(\ref{eqn:logchi}) in terms of $\lambda$,
we find
\begin{equation}
  \langle\langle Q^k(t) \rangle\rangle = M\sum_j
    \left(T(1-T){d \over dT}\right)^{k-1}T|_{T=T_j} \, .
  \label{eqn:kthcumulant}
\end{equation}
We note that both $\log \chi(\lambda)$ and
$\langle\langle Q^k(t) \rangle\rangle$ are linear statistics of $T_j$'s.

Current fluctuations are determined by distribution of
transmission coefficients and the distribution varies
from sample to sample even though samples have
the same macroscopic parameters.
Therefore in principle each sample exhibits distinctive current
fluctuations.
However according to the random matrix theory of disordered conductors,
in the metallic regime($1 \! \ll \! g \! \ll \! N$) where $N$ is
the number of channels,
the distribution approaches a universal one\cite{Stone}.
This result provides a motivation to approximate the sample-dependent
distribution by the universal one.
To exploit the universal distribution,
we introduce  new variables $\nu_j$'s and the density function $D(\nu)$
defined by $T_j=1/\cosh^2 \nu_j$ and $D(\nu)d\nu=D(T)dT$,
where $D(T)$ is the density function of $T_j$'s.
According to Ref.~\cite{Stone}, $D(\nu)$ is uniform over a wide range
of $\nu$,
\begin{equation}
  D(\nu)=g \,\,\mbox{for }\nu<\nu_c.
  \label{eqn:conversion}
\end{equation}

We combine Eq.~(\ref{eqn:logchi})
with the universal distribution to obtain
\begin{equation}
  \overline{ \log \chi(\lambda)}=Q_0\int_{0}^{\infty} d\nu
  \log\Biggl({e^{i\lambda}-1 \over \cosh^{2}\nu}+1 \Biggr),
  \label{eqn:integration}
\end{equation}
where $Q_0=gM$.
In Eq.~(\ref{eqn:integration}) the upper limit $\nu_c$
is replaced by infinity, which is valid in
the metallic regime because for large $\nu$, the integrand is
exponentially small. The evaluation of the integral then leads to
Eq.~(\ref{eqn:result}). We note that because $\log\chi(\lambda)$ is
a linear statistic, the universal distribution approximation
is equivalent to taking an average over samples.

Cumulants are useful in understanding features of the probability
distribution. By using the formula Eq.~(\ref{eqn:formulacumulant}),
we obtain sample averaged cumulants
\begin{equation}
  \begin{array}{rclrcl}
    \overline{ \langle\langle Q(t) \rangle\rangle} & = &
      Q_0 ,  &
    \overline{ \langle\langle Q^2(t) \rangle\rangle} & = &
      {1 \over 3}Q_0 , \\
    \overline{ \langle\langle Q^3(t) \rangle\rangle} & = &
      {1 \over 15}Q_0 , &
    \overline{ \langle\langle Q^4(t) \rangle\rangle} & = &
      -{1 \over 105} Q_0 , \\
    \overline{ \langle\langle Q^5(t) \rangle\rangle} & = &
      -{1 \over 105}Q_0 , &
    \overline{ \langle\langle Q^6(t) \rangle\rangle} & = &
      {1 \over 231}Q_0 , \\
    \overline{ \langle\langle Q^7(t) \rangle\rangle} & = &
      {27 \over 5005}Q_0 , &
    \overline{ \langle\langle Q^8(t) \rangle\rangle} & = &
      -{3 \over 715} Q_0 , \\
    \overline{ \langle\langle Q^9(t) \rangle\rangle} & = &
      -{233 \over 36465}Q_0 , &
    \overline{ \langle\langle Q^{10}(t) \rangle\rangle} & = &
      {6823 \over 969969}Q_0 ,  \cdots \, .
  \end{array}
  \label{eqn:cumulants}
\end{equation}
The first cumulant is trivial. It is
just a definition of $G$ and it shows where the peak of $P(Q(t))$ is.
The second cumulant measures the width square
of the peak. It is also directly related to the noise power
$P=\int \! dt \, \langle\langle I(0)I(t) \rangle\rangle$,
a widely used measure of noise magnitude,  by
$\langle\langle Q^2(t) \rangle\rangle=tP$ for large $t$,
and its ensemble average is one-third of the classical value $Q_0$,
as first pointed out by Beenakker and B\"uttiker\cite{Beenakker}.
The third and the fourth cumulants are measures of skewness and sharpness
of the peak, respectively and they are related to $3$ and $4$-point
current current correlation functions by similar relations.
We note that all cumulants are proportional to $Q_0$ and that
for $Q_0 \gg 1$, $\overline{ \langle\langle Q(t) \rangle\rangle}^k \gg
\overline{ \langle\langle Q^k(t) \rangle\rangle}$. Therefore the peak
of the distribution $P(Q(t))$ is Gaussian for large conductance limit
or long time limit. This result is quite expected from the central
limit theorem. Now to see the tails of $P(Q(t))$, we study higher order
cumulants. From Eq.~(\ref{eqn:integration}), we obtain a general formula
for the ensemble averaged $k$-th order cumulants
\begin{equation}
  \overline{ \langle\langle Q^k(t) \rangle\rangle}=-i^k{Q_0 \over 4}
    \int^\infty_{-\infty} dx{1 \over \sqrt{1+e^{-x}}}
    \int^\infty_{-\infty}
    dq \, e^{-iqx} {q^{k-1} \over \sinh (\pi q-i0+)},
  \label{eqn:avekthcumulant}
\end{equation}
and by using the steepest descent method twice,
we obtain the asymptotics for large $k$,
\begin{equation}
  \overline{ \langle\langle Q^k(t) \rangle\rangle}\sim
    {Q_0 \over (2\pi)^{k-1}}{(k-1)! \over \sqrt k} \left\{
    \begin{array}{cc}
      (-1)^{k+2 \over 2} & \mbox{for even $k$}, \\
      (-1)^{k+1 \over 2} & \mbox{for odd $k$}.
    \end{array} \right\}
  \label{eqn:asymptotics}
\end{equation}
The high order cumulants diverge as factorial, which suggests
that at the tails, $P(Q(t))$ is different from both
Gaussian distribution and Poisson
distribution which describes the classical current fluctuations.
In comparison,
$\overline{ \langle\langle Q^k(t) \rangle\rangle}=0$ for $k\ge 2$ for
Gaussian distribution and
$\overline{ \langle\langle Q^k(t) \rangle\rangle}=Q_0$ for $k\ge 1$ for
Poisson distribution.

It is known that in the presence of time reversal symmetry,
there are order $M$ corrections to
$\overline{\langle\langle Q(t) \rangle\rangle}$ and
$\overline{\langle\langle Q^2(t) \rangle\rangle}$
due to weak localization\cite{de Jong}, and
it is legitimate to expect the same kind of corrections to
higher order cumulants. However, because we are interested
in the metallic regime, these corrections are small by a factor $g$
and we will ignore them.

A proper next step is to estimate the magnitude of sample to
sample variations of $P(Q(t))$. Here instead of $\log \chi(\lambda)$,
we examine variance of $\langle\langle Q^k(t) \rangle\rangle$
to see the variations
of $P(Q(t))$. $\langle\langle Q^k(t) \rangle\rangle$ is
a linear statistic and the general formula for the variance
of a linear statistic $A=\sum_{j}a(T_j)$ is
obtained recently by Beenakker and Rejaei\cite{Rejaei}, and Chalker and
Mec\^{e}do\cite{Chalker},
\begin{eqnarray}
  \mbox{var }(A)={1 \over \beta \pi^2}\int_{0}^{\infty}dk
    {k\tilde{a}^2(k) \over 1+\coth({1 \over 2}\pi k)} \, ,
  \label{eqn:varformula} \\
  \tilde{a}(k)=2\int_{0}^{\infty} d\nu \, a\left( {1 \over \cosh^2 \nu}\right)
    \cos k\nu \, ,
  \label{eqn:varformula2}
\end{eqnarray}
where $\beta$ is a symmetry constant, $1,2$, or $4$ depending on
the symmetry.
We use this formula to obtain
\begin{equation}
  \begin{array}{rclrclrcl}
    \mbox{var}\left(\langle\langle Q(t) \rangle\rangle\right) & = &
          {2 \over 15\beta}M^2 \, ,&
    \mbox{var}\left(\langle\langle Q^2(t) \rangle\rangle\right) & = &
          {46 \over 2835\beta}M^2 \, , &
    \mbox{var}\left(\langle\langle Q^3(t) \rangle\rangle\right) & = &
          {11366 \over 1447875\beta}M^2 \, , \cdots .
  \end{array}
  \label{eqn:variancecumulant}
\end{equation}
We note that for low order cumulants,
$\overline{ \langle\langle Q^k(t) \rangle\rangle}^2$ is larger than
$\mbox{var}\left(\langle\langle Q^k(t) \rangle\rangle\right)$
by at least a factor of $g^2$, which is large in the metallic regime.
Low order cumulants decide the shape of $P(Q(t))$ around the peak and
therefore the small variance of low order cumulants implies that
the peak shape shows little sample to sample variations, that is,
it is almost universal.

To see the behavior of higher order cumulants, we obtain an asymptotic
form of the variance from an approximate variance formula in
Ref.~\cite{Beenakker2},
\begin{equation}
  \mbox{var }\left(\langle\langle Q^k(t) \rangle\rangle\right) \sim
    {4(2k-1)! \over (2\pi)^{2k}\beta}M^2.
  \label{eqn:asytoticvarcum}
\end{equation}
According to Eq.~(\ref{eqn:asytoticvarcum}), for high order cumulants,
$\mbox{var}\left(\langle\langle Q^k(t) \rangle\rangle\right)$ becomes
larger than $\overline{ \langle\langle Q^k(t) \rangle\rangle}^2$ due to
its rapidly growing factorial factor, which suggests that
the tails of $P(Q(t))$ show large sample to sample variations.
We argue that this rapid growth of
$\mbox{var}\left(\langle\langle Q^k(t) \rangle\rangle\right)$
is not an artifact of the approximate variance formula used above
because it assumes stronger spectral rigidity than the formula
Eq.~(\ref{eqn:varformula},\ref{eqn:varformula2}) and it has a tendency to
slightly underestimate variances. Therefore large sample to
sample variations at the tails of $P(Q(t))$ obtained above
is not an artifact of the approximation.

In the above, we derived the shape of $P(Q(t))$ by examining
$\overline{ \log \chi(\lambda)}$ and its cumulant expansion
instead of $\log \overline{\chi(\lambda)}$, which might be
an intuitively more appropriate ensemble average
because it is directly related to $\overline{P(Q(t))}$.
However we argue that in contrast to the intuition,
$\overline{ \log \chi(\lambda)}$ is an appropriate ensemble
average for the study of current fluctuations.
One reason is that as we remarked earlier,
a $k$-point current current correlation function is linearly
related to $\langle\langle Q^k(t) \rangle\rangle$, whose
ensemble average can be obtained from
$\overline{ \log \chi(\lambda)}$ by a simple expansion.
Another reason is that as we show later,
$\log \overline{\chi(\lambda)}$ either becomes identical to
$\overline{ \log \chi(\lambda)}$ at short time limit, or
is dominated by the conductance fluctuations instead of
the current fluctuations.

Calculation of $\overline{\chi(\lambda)}$
is not simple because $\chi(\lambda)$ is not a linear statistic.
Muttalib and Chen\cite{Chen}
did this calculation recently by large $N$ limit continuum approximation
and showed that at long time limit, $\log\overline{\chi(\lambda)}$
becomes quite different from $\overline{\log \chi(\lambda)}$.
Here we present improved calculation by a perturbation method
and we believe that our calculation clarifies the reason
why two averages become quite different at long time limit.

Because $\chi(\lambda)$ is not a linear statistic,
we need joint probability distribution of transmission coefficients
to average it over ensembles. After standard variable change,
$T=1/(1+x)$, the joint probability distribution $P(\{x\})$ is
\begin{equation}
  P(\{x\})=\exp\left(\beta\sum_{a<b}V(x_a,x_b)+\beta\sum_a U(x_a)\right).
  \label{eqn:jointprob}
\end{equation}
We choose $V(x,y)=(1/2)\log(x-y)+(1/2)\log({\rm arcsinh}^2\sqrt{x}
  -{\rm arcsinh}^2\sqrt{y})$
and $U(x)=g\,{\rm arcsinh}^2(\sqrt{x})$ based on the exact calculation
of the joint probability distribution function for $\beta=2$
by Beenakker and Rejaei\cite{Rejaei}.
Then,
\begin{equation}
  \begin{array}{c}
    \overline{\chi(\lambda)}={Z_M \over Z_0} \, ,
      \label{eqn:avechifor} \\
    Z_M=\int \prod_a dx_a
      \exp\left(\beta\sum_{a<b}V(x_a,x_b)+\beta\sum_a U(x_a)
        +\sum_a M\log{x_a+e^{i\lambda} \over x_a+1}\right)
      \, , \\
    Z_0=\int \prod_a dx_a
      \exp\left(\beta\sum_{a<b}V(x_a,x_b)+\beta\sum_a U(x_a)\right)
      \, .
  \end{array}
\end{equation}
By expanding $\log \overline{\chi(\lambda)}$ in terms of $M$,
we find
\begin{equation}
  \log \overline{\chi(\lambda)}=\overline{\log\chi(\lambda)}
    +\mbox{var}\left(\log\chi(\lambda)\right)+O(M^3) \, ,
  \label{eqn:avechiexp}
\end{equation}
and from the formula Eq.~(\ref{eqn:varformula},\ref{eqn:varformula2}),
we obtain
\begin{equation}
  \log \overline{ \chi(\lambda)}=gM{\rm arcsinh}^2\sqrt{e^{i\lambda}-1}
    -{M^2 \over \beta}
    \left(3\log{ {\rm arcsinh} \sqrt{e^{i\lambda}-1}
       \over \sqrt{e^{i\lambda}-1}}
    +{1 \over 2}i\lambda\right)+O(M^3).
  \label{eqn:averagechi}
\end{equation}
Note that for $M \ll g$(short time limit),
$\log \overline{ \chi(\lambda)}$ reduces to
$\overline{ \log\chi(\lambda)}$.
We expand $\log \overline{ \chi(\lambda)}$ in terms of $\lambda$
to see features of $\overline{ P(Q(t))}$:
\begin{equation}
  \log \overline{ \chi(\lambda)}=gM(i\lambda)+
     \left({g \over 3}M+{2 \over 15\beta}M^2\right)
        {(i\lambda)^2 \over 2!}+
     \left({g \over 15}M+{2 \over 315\beta}M^2+O(M^3)\right)
        {(i\lambda)^3 \over 3!}+
     \cdots .
  \label{equn:avechiseries}
\end{equation}
The first expansion coefficient shows that
$\overline{\langle Q(t) \rangle}=Q_0=gM$, which is trivial.
The second expansion coefficient,
$\overline{ \langle Q^2(t) \rangle}-\overline{ \langle Q(t) \rangle}^2
 =\overline{ \langle\langle Q^2(t) \rangle\rangle}+
  \mbox{var}(\langle Q(t) \rangle)
 =Q_0/3+(2 / 15\beta)M^2$ indicates that the peak width
of $\overline{P(Q(t))}$ has two contributions. The first contribution
is related to the noise power, and the second one to
the universal conductance fluctuations because
$\mbox{var}(\langle Q(t) \rangle)$
is proportional to the variance of the conductance.(The factor
$2/15 \beta$ is precisely the variance of the dimensionless conductance.)
Note that as $t \rightarrow \infty$, the second contribution becomes
dominant over the first one.
It can be shown that the $k$-th order expansion
coefficient contains $k$ different contributions and
at long time limit, the most dominant contribution, which is proportional
to $M^k$, is related to  the $k$-th cumulant of the conductance
fluctuations.
{}From this analysis
we see that the behavior of $\overline{\chi(\lambda)}$ for large $t$
is governed by the conductance fluctuations instead of
the current fluctuations.

As a short remark, we report that the continuum approximation
calculation of $\log \overline{\chi(\lambda)}$, as suggested by
Muttalib and Chen\cite{Chen}, produces the same result as
Eq.~(\ref{eqn:averagechi}) up to $M^2$ order.

In summary, we examine the counting statistics of charge
to study the low temperature current fluctuations.
By calculating the characteristic function of the probability
distribution $P(Q(t))$, we find that $P(Q(t))$ has a Gaussian
peak at $Q_0$ with $\langle\langle Q(t) \rangle\rangle=Q_0/3$
and we also find that
the tails of $P(Q(t))$ are different from the tails of
Gaussian and classical Poisson distribution.
By studying the variances of the cumulants, we establish
that even though the peak location of $P(Q(t))$ varies from sample
to sample due to universal conductance fluctuations, the peak shape
of $P(Q(t))$ is universal in the metallic regime, and that
the sample to sample variations show up only
at the tails of $P(Q(t))$.


\begin{references}

\bibitem{Nyquist} J. B. Johnson, Phys. Rev. {\bf 29}, 367 (1927) and
  H. Nyquist, Phys. Rev. {\bf 32}, 110 (1928).
\bibitem{Landauer} R. Landauer, in {\it Localization, Interaction
  and Transport Phenomena}, eds. B. Kramer, G. Bergmann and
  Y. Bruynsraede (Springer, Heidelberg, 1985) Vol {\bf 61}.
\bibitem{Lesovik} G. B. Lesovik, Pis'ma Zh. Eksp. Teor. Fiz.
  {\bf 49}, 513 (1989) [JETP Lett. {\bf 49}, 592 (1989)].
\bibitem{Yurke} B. Yurke and G. P. Kochanski, Phys. Rev. B {\bf 41},
  8184 (1990).
\bibitem{Buttiker} M. B\"uttiker, Phys. Rev. Lett. {\bf 65}, 2901
  (1990).
\bibitem{Stone} For a review see: A. D. Stone, P. A. Mello,
  K. A. Muttalib, and J.-L. Pichard, in {\it Mesoscopic Phenomena
  in Solids}, Edited by B. L. Al'tshuler, P. A. Lee, and R. A. Webb
  (North-Holland, Amsterdam, 1991).
\bibitem{Beenakker} C. W. J. Beenakker and M. B\"uttiker, Phys.
  Rev. B {\bf 46}, 1889 (1992).
\bibitem{LevitovBinomial} L. S. Levitov and G. B. Lesovik, JETP Lett.
  {\bf 58} (3), 230 (1993).
\bibitem{Mello} P. A. Mello, P. Pereyra, and N. Kumar, Ann. Phys.
  {\bf 181}, 290 (1988).
\bibitem{comment} The spin degeneracy is ignored. To include
  the degeneracy, $M$ has to be multiplied by $2$. Also
  the positivity of $M$ is assumed. If $M$ is negative,
  Eq.~(\ref{eqn:binomial}) has to be complex conjugated with $M$
  replaced by its absolute value.
\bibitem{de Jong} M. J. M. de Jong and C. W. J. Beenakker,
  Phy. Rev. B {\bf 46}, 13400 (1992).
\bibitem{Rejaei} C. W. J. Beenakker and B. Rejaei, Phys. Rev. Lett.
  {\bf 71}, 3689 (1993).
\bibitem{Chalker} J. T. Chalker and A. M. S. Mac\^{e}do,
  Phys. Rev. Lett. {\bf 71}, 3693 (1993).
\bibitem{Beenakker2} C. W. J. Beenakker, Phys. Rev. Lett. {\bf 70},
  1155 (1993).
\bibitem{Chen} K. A. Muttalib and Y. Chen, preprint
  (cond-mat/papers/9405068@babbage.sissa.it).
\end{references}
\end{document}